\documentclass[runningheads]{llncs}
\usepackage{amsmath}
\usepackage[T1]{fontenc}
\usepackage{graphicx}
\begin{document}
\title{Developing a Multi-Agent System to Generate Next Generation Science Assessments with Evidence-Centered Design}
\titlerunning{Multi-Agent Generation of NGSS Assessments with ECD}

%
%
\author{Yaxuan Yang\inst{1,2} \and
Jongchan Park\inst{1} \and
Yifan Zhou\inst{1,3} \and
Xiaoming Zhai\inst{1,4}}
\authorrunning{Yang et al.}
%
\institute{AI4STEM Education Center, University of Georgia, Athens, GA 30602, USA \and
Department of Educational Psychology, University of Georgia, Athens, GA 30602, USA \and
School of Computing, University of Georgia, Athens, GA 30602, USA \and
Department of Mathematics, Science, and Social Studies Education, University of Georgia, Athens, GA 30602, USA
\\
}
\maketitle              
\begin{abstract}
Contemporary science education reforms such as the Next Generation Science Standards (NGSS) demand assessments that capture students’ ability to use science knowledge to solve problems and design solutions. To elicit such higher-order ability, educators need performance-based assessments, which are challenging to develop. One solution that has been broadly adopted is Evidence-Centered Design (ECD), which emphasizes interconnected models of the learner, evidence, and tasks. Although ECD provides a framework to safeguard assessment validity, its implementation requires diverse expertise (e.g., content and assessment), which is both costly and labor-intensive. To address this challenge, this study integrated ECD into a Multi-Agent System (MAS) to automatically generate NGSS-aligned assessment items by combining multiple large language models with complementary expertise to automate multi-stage item generation workflows typically handled by human experts. We evaluated the quality of MAS-generated items and compared them to human-developed items across multiple dimensions of assessment design. Results showed that MAS-generated items have overall comparable quality to human-developed items regarding alignment with NGSS three-dimensional standards and cognitive demands. Divergent patterns also emerged: MAS-generated items demonstrated a distinct strength in inclusivity, while exhibiting limitations in clarity, conciseness and multimodal design. MAS- and human-developed items both showed weaknesses in evidence collectability and student interest-alignment. These findings suggest that integrating ECD into MAS can support scalable and standards-aligned assessment design, while human expertise is still essential.

\keywords{Multi Agent Systems  \and Evidence-Centered Design \and Automated Assessment Design.}
\end{abstract}
\section{Introduction}
Assessment practices have evolved with changes in technology~\cite{meylani2024}. That is, assessment has become more personalized and interactive, such as computer-based testing and adaptive assessments. At the same time, technology has made the process of assessment design with higher efficiency. Traditional assessment design is labor-intensive and time-consuming, which is difficult to meet the increasing demand of scaling up~\cite{prasetyo2020}. Meanwhile, technologies such as artificial intelligence (AI) facilitated automated item generation (AIG), making it easier to produce large numbers of items with programming algorithms~\cite{alves2010}. Moreover, researchers reported that AIG is able to produce high-quality items that are similar to experts’ level~\cite{prasetyo2020}. 

Besides the advancement of technology, recent education reforms also emphasize the importance of aligning assessments with national science standards~\cite{penuel2015}. For example, the Next Generation Science Standards (NGSS) provide an opportunity to transform science education for all students within a three-dimensional model of learning, including disciplinary core ideas (DCI), science and engineering practices (SEP), and crosscutting concepts (CCC;~\cite{penuel2015}). Within the NGSS framework, higher-order -ability refers to students’ capacity for knowledge-in-use learning, in which DCI are applied through SEP and CCC to construct explanations, analyze data, develop or use models, and design solutions. As for this, there is a need to develop a framework that can relate learning goals with evidence of student understanding. 

Evidence-Centered Design (ECD) is a layered architecture that meets the requirements of NGSS-aligned assessments. It claims that assessment design has to be clear about what they want to claim, how to get the evidence, and how the assessment elicits the evidence~\cite{mislevy2003}. However, implementing ECD remains largely labor-intensive, since experts have to manually make inputs for each step~\cite{mislevy2012}. This limitation highlights the need for a computational framework to implement ECD principles in an automated process. 

As a computational framework, Large Language Models (LLMs) bring transforms for test development and have been applied in educational assessment~\cite{lee2025aig}. Unlike traditional assessment design, LLM can work on multi-tasks, including generation, classification, and logical reasoning. Based on LLMs, recent studies have explored more opportunities and proposed multi-agent frameworks in educational contexts~\cite{chu2025}. A multi-agent framework is a system architecture that coordinates multiple role-specialized agents, often powered by an LLM, to decompose complex work into subtasks and integrate their outputs to support automated assessment design~\cite{lee2025aig}. While LLM-based agents have been applied to educational tasks such as assessment, adaptive learning, and feedback generation~\cite{chu2025}, researchers have limited understanding on how theoretical framework such as ECD can be integrated into LLM-based agents to ensure the validity and follow the learning standards. 

This study develops and evaluates an ECD-based multi-agent system that automates assessment design to generate NGSS-aligned assessment items. The evaluation was guided by two research questions: (1) What is the quality of the AI-generated NGSS-aligned assessments compared to human expert developed? (2) What are the differences between AI and human experts in writing NGSS-aligned assessment items? 
\section{Literature Review}
\subsection{Next Generation Science Assessments}
Next Generation Science Assessment (NGSA) is a classroom formative assessment initiative aligned with the Next Generation Science Standards~\cite{harris2019}. NGSA was developed to support assessment designs that move beyond a traditional emphasis on isolated content recall toward knowledge-in-use, in which students demonstrate integrated use of science and engineering practices, disciplinary core ideas, and crosscutting concepts as they work toward grade-band performance expectations~\cite{zhai2024}. Designing NGSS-aligned assessment items under this goal requires a systematic development approach. Designers should analyze the target domain to identify focal performance expectations and their associated three-dimensional components, articulate the evidence needed to demonstrate the relevant knowledge, skills, and abilities, and specify task features that can elicit that evidence in interpretable ways. To structure and support this complex design work, NGSA adopts evidence-centered design (ECD) as its guiding development framework~\cite{harris2019}. 

\subsection{Evidence-Centered Design (ECD)}
ECD is a systematic evaluation design framework that is used to construct the logic chain from observable behaviors to cognitive reference~\cite{riconscente2015}. It highlights that assessment is a process to infer students’ skills and knowledge based on their observable behaviors. According to Mislevy et al.~\cite{mislevy2003}, this process can be divided into three core layers. The first layer is domain model, defining the targeted ability or knowledge that wants to assess. The second layer is evidence model, defining the specific observable behaviors as evidence. The last layer is task model, defining specific task scenarios to assess in order to elicit the evidence. These three layers form the claims-evidence-tasks reasoning chain, where tasks elicit evidence, evidence supports inference, and finally verifies the traceability and interpretability~\cite{mislevy2006}. 

Researchers showed that ECD framework has several advantages~\cite{mislevy2012,riconscente2015}. It enhances the assessment validity by ensuring the reasoning path and the consistency between assessment design and interpretive reasoning. Moreover, ECD framework supports the reusability and scalability, making tasks and scoring logic reusable across projects. It also promotes communication efficiency, since it provides a shared language that connects different groups of experts, such as content, measurement, and technical specialists. 

However, ECD has its limitations for large-scale assessment development~\cite{mislevy2012}. It relies heavily on human experts, especially for domain modeling and evidence specification. Experts have to define targeted knowledge or skills, observable behaviors, and design the task scenario and score rubrics, making it difficult for scalable assessment. These resource demands create a strong need for AI-assisted workflows that can accelerate development process while preserving ECD’s claims-evidence-task chains.

\subsection{Automated Item Generation (AIG)}
AIG and LLMs provide the opportunity to apply ECD to scalable level. AIG is a computer-based method that can produce scalable test items automatically based on structured templates or item models~\cite{gierl2013}. Commonly, AIG can be categorized into automatic and semi-automatic models~\cite{prasetyo2020}. Automatic models completely rely on algorithms to generate items without any human intervention, while semi-automatic models still need experts’ input to define stems, providing more flexibility across disciplines. Compared to traditional item design, AIG largely improved efficiency, consistency and reduced cost~\cite{gierl2013}.

Despite its advantages, researchers argued that AIG does not have a strong measurement foundation~\cite{tan2024}. AIG has been largely engineering-oriented, with comparatively limited attention to the validity, reliability, and pedagogical alignment of generated items. In their review study, Tan et al.~\cite{tan2024} summarized that few studies explicitly specify the targeted cognitive skills or examine how generated items align with the learning goals, which are the core of educational assessment theories like ECD. Additionally, many AIG approaches incorporate multimodal elements such as images, tables, and graphs, but the coherence among these multimodal representations has not been examined comprehensively in prior AIG research~\cite{tan2024}. Researchers have noted that poorly coordinated multimodal representations can increase ambiguity and cognitive load for students, particularly in assessment contexts~\cite{mayer2002}.
\subsection{Multi-Agent System (MAS)}
Agent refers to the LLMs powered autonomous entity that can achieve certain goals through planning, reasoning, and executing tasks~\cite{guo2024}. MAS is an expanded idea of agent. In a MAS, multiple LLM-based agents are assigned specific roles~\cite{li2024mas}. They work together by communicating, collaborating, and sharing the reasoning steps to do a complex task that is difficult for a single model. In educational context, MAS has been applied to adaptive learning, feedback generation and evaluation, where multiple LLM-based agents play different roles to simulate human-like reasoning and dialogue~\cite{chu2025}. 

MAS has the potential to mitigate limitations of current AIG. Recent works extended the MAS to AIG, proposing a new framework called LLM-based Multi-Agent AIG (LM-AIG;~\cite{lee2025aig}). It basically assigns roles to each agent, such as item writer, content reviewer, language and bias reviewer, and lets them work collaboratively to do the assessment design. This framework used a human-in-the-loop mechanism, highlighting that the generated items are similar to human experts’ level in clarity and interpretability but need human reviewer in construct validity.

\section{Methodology}
\subsection{Overview}
Building on the recent works on LLM-based MAS, this study proposes a new framework that integrates principles of ECD into a multi-agent framework to develop NGSS-style assessment tasks automatically, called ECD-based MAS. Based on the principles of ECD, this study summarizes ECD into the following steps. 
\subsection{Function of Each Agent}
\begin{figure}
\includegraphics[width=\textwidth]{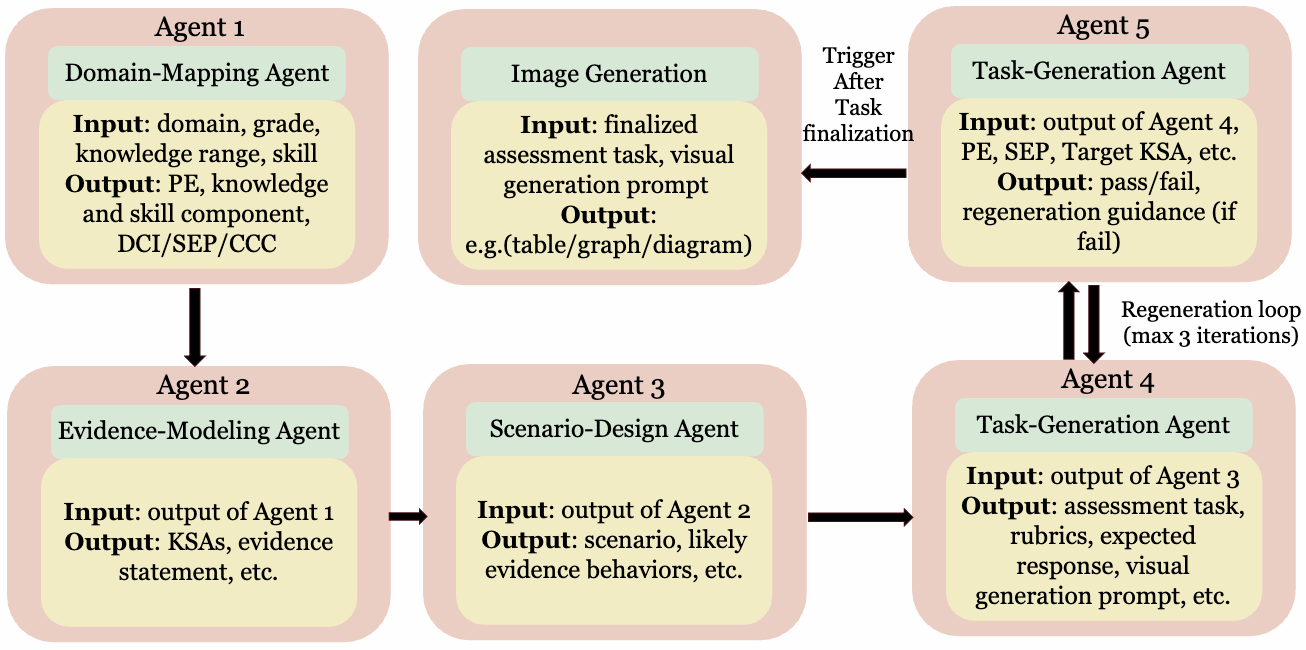}
\caption{Architecture of the ECD-based MAS for NGSS-aligned assessment design, with an evaluation-regeneration loop and an image generation step.} \label{fig2}
\end{figure}

There are five agents in total (see Fig.~\ref{fig2}). Agent 1 works as the domain modelling agent. It identifies the performance expectation (PE) based on the given domain, grade band, and knowledge range. It outputs a structured representation that includes the DCI, SEP, CCC, and concise knowledge and skill components to anchor subsequent design stages. Agent 2 serves as the evidence modelling agent. It expands the PE into explicit Knowledge, Skills, and Abilities (KSAs) and corresponding evidence statements. It is asked to describe observable behaviors, partial understandings, and misconceptions. Agent 3 works as the scenario design agent. It generates daily-life scenarios that can elicit the evidence behaviors specified by the second agent. Moreover, it specifies the context, task conditions, and the likely evidence behaviors expected to elicit in performance situations. Agent 4 acts as the item generation agent. It transforms the scenarios into assessment tasks, including task prompt, expected student response, scoring rubric, task features, fairness considerations, with explicit constraint. Agent 5 works as an independent quality evaluation agent, assessing whether the generated assessment tasks meet predefined NGSS quality criteria, including three-dimensional alignment, cognitive demands, language clarity, and cross-agent consistency. This agent does a pass/fail judgment, providing a final quality control step in the multi-agent pipeline. 

\subsection{Implementation Details}
In this ECD-based MAS, each agent is implemented through role-specific prompts, defining the responsibility scope explicitly. To ensure basic quality control while avoiding unbounded iteration, the evaluation-regeneration loop sets three as the maximum number of retries. If the task does not meet the quality criteria after three attempts, the system retains the final generated version as the output. After the assessment task is finalized, the system generates a corresponding visual representation using Gemini Imagen 4.0. To improve the image quality, the process is based on a structured prompt combining detailed scene description and visual generation directives, guiding the model toward producing contextually appropriate visuals aligned with the assessment scenario. 

Splitting the whole workflow into multiple agents with clearly defined responsibilities is motivated by both theoretical consideration and practical needs. To ensure coherence and consistency, we made a series of structural constraints to the multi-agent system. First, the system adopts a unidirectional, inheritance-based information flow. Each agent only uses the structured outputs from the upstream agents, preventing information backflow. Second, the system restricts the source of KSAs and evidence statement strictly to the aligned NGSS knowledge and skill dimensions, preventing construct drift that may be introduced by additional constructs or implicit abilities. Moreover, during the task generation step, evidence is asked to be instantiated as observable data representations, such as tables and graphs, and use as the direct basis for student reasoning. This constraint enforces an evidence-first assessment principle. 

All agents are executed by the large language model (GPT-4o-mini). To ensure the traceability across different agents, this multi-agent system uses a JSON-based input-output design, where each agent produces a structured JSON output that can be passed to the next agent. Strict output schemas and prompt constraints are applied to improve controllability. This system also relies on a sequential pipeline with iterative refinement, in which each agent works based on the output from the previous agent, reflecting the claims-evidence-task structure of ECD framework. 

\subsection{Evaluation}
To evaluate the performance of the ECD-based MAS, we conducted a comparative analysis of MAS-generated and human-developed NGSA items using a multidimensional rubric grounded in assessment quality, engagement, and multimedia design. 

\subsubsection{Item pool}
We assembled two parallel item pools aligned with the same NGSS performance expectations. The existing NGSA items consisted of 15 life science items targeting MS-LS1 and MS-LS2 and 15 physical science items targeting MS-PS1, MS-PS2, and MS-PS3 ($n = 30$ total). The MAS-generated items were produced for the same performance expectations (15 life science, 15 physcial science, $n = 30$). 
\subsubsection{Evaluation rubric development}
We developed an NGSA evaluation rubric by adapting three established frameworks: a high-quality assessment framework using Depth of Knowledge (DOK) to index cognitive demand ~\cite{darling2013}, a performance-assessment engagement framework using the relevance dimension ~\cite{taylor2016}, and multimedia learning theory using the coherence, signaling, and spatial-contiguity principles ~\cite{mayer2002}. We selected these constructs because they directly support item-level judgments; framework elements focused on system-level purposes (e.g., international benchmarking) or task administration (e.g., autonomy and collaboration) were excluded. We also added NGSS-specific criteria for explicit alignment with target performance expectations (integrating DCI, SEP, and CCC) and for clarity, conciseness, and appropriateness of language.  

The final rubric included five overarching criteria and 14 components: (a) NGSS 3D alignment (five components: four dichotomous [Satisfies/Does Not Satisfy] and one three-point scale), (b) cognitive demand (one component: DOK 3-4, 2, or 1), (c) student engagement (two components), (d) clarity, conciseness, and appropriateness of language (three components), and (e) multimodal representation (three components). Except for the four dichotomous alignment components and the DOK rating, all components used a three-point scale (Evident, Partially Evident, or Not Evident).

\subsubsection{Rater training and scoring}
Two science education experts—one with a doctorate and one with a master’s degree—with experience in NGSS curriculum and assessment design served as raters. Inter-rater reliability was estimated using Gwet’s~\cite{gwet2014} chance-corrected agreement coefficients: AC$_1$, for four dichotomous components, and quadratic-weighted AC$_2$ for ordinal components. In Round~1 ($n = 10$ items randomly selected), agreement was unanimous for the dichotomous components and DOK ($\text{AC}_{1} = 1.00$). Overall agreement across nine three-point components was strong ($\text{AC}_{2} = 0.90$, 95\% CI [$0.85$, $0.96$]), but the Evidence Collectivity component showed lower reliability ($\text{AC}_{2} = 0.68$, 95\% CI [$0.46$, $0.90$]), requiring rule refinement between the two raters. The raters met to resolve discrepancies, clarify rubric language, and refine shared decision rules. In Round 2 ($n = 20$ items randomly selected), the raters independently scored an additional 20 items (10 MAS-generated and 10 NGSA items), achieving perfect agreement ($\text{AC}_{1} = 1.00$, $\text{AC}_{2} = 1.00$).
 
\subsubsection{Main scoring and analysis plan}
After training, the two raters independently scored the remaining 30 items (15 MAS-generated and 15 NGSA items) using the rubric. Disagreements were resolved through discussion, and consensus ratings were used in subsequent analyses. 

For Research Question 1, we summarized rubric ratings by item source for each component using item-level descriptive statistics. For Research Question 2, we conducted a complementary qualitative analysis to identify recurring patterns in item features that explained observed differences between MAS-generated and NGSA items. The first and second authors performed an inductive thematic analysis of item prompts and associated visual representations, using rubric-based scores as analytic context. The analysis followed the six-step procedure described by~\cite{kiger2020}: familiarization, initial coding, searching for themes, reviewing themes, defining and naming themes, and producing the report.

\section{Results}
\subsection{Quality of AI-generated NGSS-aligned Assessments}
At an overall level, the scoring result showed that MAS-generated items have comparable quality to human-developed NGSA items in terms of alignment with NGSS three-dimensional standards and cognitive demands (see Table~\ref{tab:rubric}). For criterion 1, both NGSA and MAS met the four components, including PE, SEP, DCI, and CCC, for all the 30 items. This result indicated that both groups can clearly target the expected performance and use related core concepts and practices. In terms of evidence collectability, both groups included items that did not meet the standards. Compared to NGSA items, MAS had higher proportions in having “partially evident” or “not evident” items, indicating a difference in eliciting evidence for students. In terms of cognitive demands in criterion 2, both groups mainly stayed at the level of DOK 3-4. Most of the MAS-generated and NGSA items were scored as requiring strategic or extended thinking, with few items at DOK 2 and none at DOK 1, suggesting comparable distributions of cognitive demands across the two groups.  

\begin{table}[!t]
\caption{Distribution of rubric ratings for NGSS-aligned assessment items generated by human experts and the MAS.}
\label{tab:rubric}
\centering
\begin{tabular*}{\textwidth}{@{\extracolsep{\fill}} p{3.2cm} p{4.6cm} p{2.2cm} p{2.2cm}}
\hline
\textbf{Criterion} & \textbf{Component} & \textbf{Human (counts)} & \textbf{MAS (counts)} \\
\hline
C1. NGSS 3D Alignment 
& Performance Expectation & 30 / 0 & 30 / 0 \\
& Science and Engineering Practice & 30 / 0 & 30 / 0 \\
& Disciplinary Core Ideas & 30 / 0 & 30 / 0 \\
& Cross-Cutting Concepts & 30 / 0 & 30 / 0 \\
& Evidence Collectability & 25 / 5 / 0 & 16 / 8 / 6 \\
\hline
C2. Cognitive Demand
& Levels 3--4 & 29 & 30 \\
& Level 2 & 1 & 0 \\
& Level 1 & 0 & 0 \\
\hline
C3. Engagement
& Relevance -- Interest & 16 / 14 / 0 & 6 / 24 / 0 \\
& Relevance -- Inclusivity & 27 / 3 / 0 & 30 / 0 / 0 \\
\hline
C4. Language
& Clarity & 29 / 1 / 0 & 2 / 25 / 3 \\
& Conciseness & 28 / 2 / 0 & 20 / 10 / 0 \\
& Appropriateness & 30 / 0 / 0 & 30 / 0 / 0 \\
\hline
C5. Multimodal Design
& Coherence & 30 / 0 / 0 & 2 / 13 / 15 \\
& Signaling & 27 / 3 / 0 & 9 / 16 / 5 \\
& Spatial contiguity & 30 / 0 / 0 & 24 / 6 / 0 \\
\hline
\end{tabular*}

\footnotesize
\textit{Note.} Values represent item counts; ratings are reported as
Satisfies / Does Not Satisfy (dichotomous) and
Evident / Partially Evident / Not Evident (three-point scale), where applicable.
Each condition includes 30 items.
\end{table}

Beyond overall quality, additional descriptive results are reported. In criterion 3, MAS-generated items consistently met the inclusivity criterion, with all items rated as evident, while NGSA items had a small number of partially evident cases. In terms of the interest, MAS showed more partially evident items than human. Criterion 4 showed both similarities and differences. Both MAS and human consistently met the criterion for language appropriateness, with no items rated as partially evident or not evident. However, there was a slight difference in conciseness and significant difference in clarity, where most of the MAS-generated items were partially evident and some of them were not evident. In criterion 5, more differences emerged in multimodal design, especially in coherence and signaling, where MAS-generated items showed a higher proportion of partially evident or not evident ratings compared to NGSA items. 

\subsection{Differences between AI and Human Experts in Writing NGSS-aligned Assessments}
Before examining differences, we note several consistent patterns across MAS-generated and human-developed NGSA items. Overall, both sources produced NGSS-aligned performance tasks that clearly targeted the intended performance expectations, with most items requiring higher-order cognitive ability. Additionally, both sources generally used grade-appropriate language. At the same time, the qualitative review revealed recurring contrasts that are elaborated in the themes below, including item prompt-level features that contributed to evidence collectability, differences in context design that reflected trade-offs between inclusivity and lived experience, and challenges that were more characteristic of MAS-generated items, particularly in linguistic clarity and multimodal coherence. 

\subsubsection{Theme 1: Evidence Collectability Hinged on Prompt Alignment and Completeness}
Evidence collectability was often compromised when prompts failed to elicit the work product implied by the targeted performance expectation or were incomplete. One recurring pattern across both sources involved a mismatch between actions dictated in the associated performance expectation and the prompt actions. For example, a MAS-generated item states, “Analyze the provided model and data to explain how food molecules undergo chemical reactions during fermentation and describe the rearrangement of these molecules.” (MS-LS1-7, biological transformations), even when the performance expectation emphasized developing a model rather than using a provided model; this mismatched prompt action changes what students are asked to produce and can reduce the likelihood that students generate the intended evidence needed for scoring. While this mismatch appeared in both NGSA and MAS-generated items, it was more frequent in MAS-generated items. MAS-generated items exhibited a distinct prompt-completeness issue, such as directing students to “answer the following questions” without providing the questions, leaving the expected evidence underspecified. 

\subsubsection{Theme 2: Contrasting Patterns in Context Design}
MAS-generated and NGSA items showed contrasting patterns in contextualizing scientific phenomena. MAS-generated items more consistently used inclusive, non-deficit language and generally avoided presuming particular home environments or resources. However, they often described phenomena in generic scientific terms rather than grounding them in concrete everyday referents (e.g., water and steam), which may limit connections to students’ lived experiences. For example, a MAS-generated item states, “monitoring the temperature and observed the motion of the particles within the liquid.” (MS-PS1-4, thermal energy). In contrast, NGSA items more often situated the phenomenon in concrete everyday situations that may support students’ sensemaking, for example, reasoning about forces using a dog puller toy in a force-interaction task. At the same time, however, some items relied on contexts that are not equally familiar across students’ living environments, potentially narrowing whose experiences are positioned as normative. For example, a NGSA item states, “Anita does some research that shows that fungi in the ground could help explain what is occurring with her lawn.” (MS-LS2-2, ecosystem interactions). Together, these patterns suggest a trade-off: MAS items tended to prioritize inclusivity through neutral framing, whereas human items more often achieved contextual specificity but occasionally relied on everyday experiences that may be unevenly accessible. 

\subsubsection{Theme 3: Linguistic Usability Reflected Different Sources of Cognitive Burden}
\begin{figure}
\includegraphics[width=\textwidth]{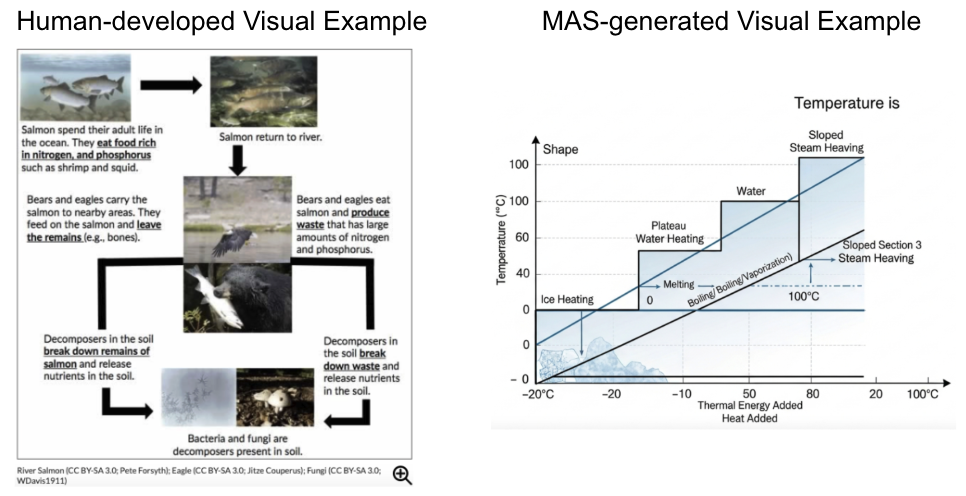}
\caption{Visual examples illustrating linguistic challenges in NGSA and MAS-generated assessment items.} \label{fig1}
\end{figure}

MAS-generated items more frequently imposed linguistic usability challenges than NGSA items, and these challenges reflected different sources of cognitive burden. In MAS-generated items, reduced usability commonly stemmed from redundancy (e.g., repeated words or labels), which made it harder for students to determine whether repetitions were intentional and meaningful or simply inadvertent (see Fig.~\ref{fig1}). Moreover, clarity was a recurring concern in MAS-generated items. While the written prompt itself was grammatically appropriate and semantically clear, key textual information embedded in image-based elements, including instructions, annotations, or captions was, in most cases, degraded, containing typographical errors, overlapping characters, non-letter symbols, or other formatting issues. In contrast, NGSA items were generally strong in linguistic usability; prompts were typically clear, grade-appropriate, and concise. When usability issues appeared in human items, they were comparatively infrequent and tended to involve occasional overelaboration, such as extended scenario narration or dense text, which increased reading load without adding values (see Fig.~\ref{fig1}). Taken together, these patterns suggest that MAS-generated items more often introduced avoidable interpretive burden, underscoring that human remains essential to refine interpretability.

\subsubsection{Theme 4: Multimodal Design as a Persistent Challenge for MAS-generated Items}
\begin{figure}
\includegraphics[width=\textwidth]{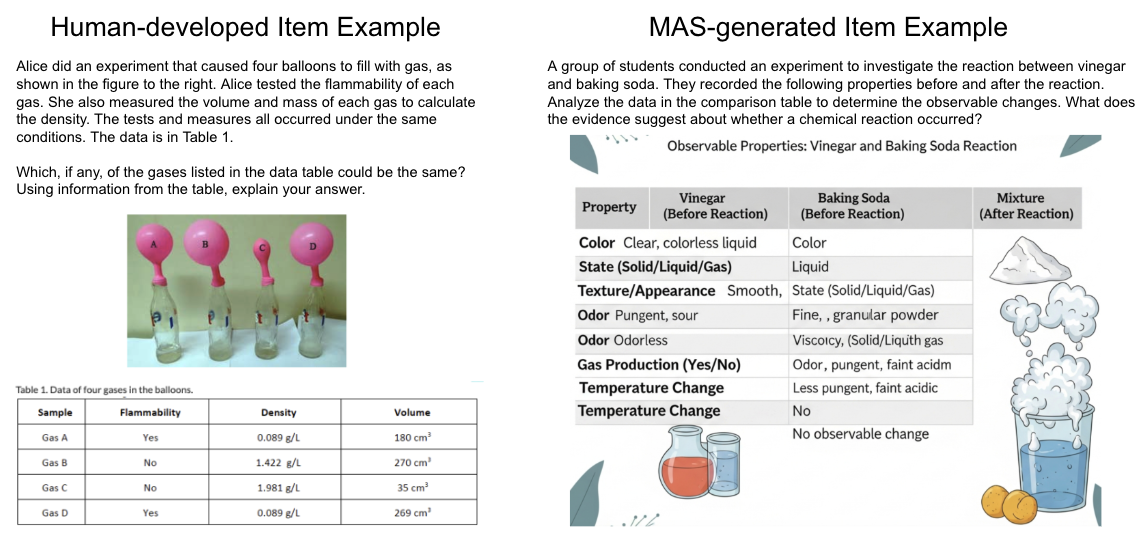}
\caption{Examples illustrating multimodal interpretability in human-developed and MAS-generated assessment items.} \label{fig3}
\end{figure}

Multimodal representation was a key divergence between MAS-generated and NGSA items. NGSA items were generally strong in integrating visuals with textual information: Tables and graphs presented interpretable information (e.g., appropriate labels and annotations) and complemented what was described in the prompt (see Fig.~\ref{fig3}). In contrast, MAS-generated items more often included visuals that were hard to interpret or poorly supported the prompt due to internal inconsistencies (e.g., graph axes, table configurations, units, or labels that did not match the quantities described in the prompt), readability problems (e.g., overcrowded visuals or overlaps between graphical elements and text) or extraneous details unrelated to the task demand (see Fig.~\ref{fig3}). Together, these patterns suggest that multimodal generation remained a key instability in MAS-produced assessment items, reinforcing the need for human review to ensure visuals function as interpretable supports for eliciting student thinking. 

\section{Discussion}
\subsection{Distinct Patterns Between MAS-Generated and NGSA Items}
These findings indicate that MAS-generated items have overall comparable quality to human-developed items but reflect different design priorities. MAS-generated items tend to use more neutral and inclusive wording and scenarios, while these items lack linguistic conciseness and clarity in multimodal design. Human-developed items are clearer but sometimes rely on everyday experiences that are not equally familiar to all students. Moreover, both groups face challenges in eliciting interpretable evidence and aligning student interests, which shows that such issues can happen for both human experts and AIG. Overall, these differences suggest that integrating ECD into MAS can support scalable and standards-aligned assessment design, while human expertise is still essential. 

\subsection{Practical Contributions to Scalable Asessment Design}
This study shows that integrating ECD into MAS can support NGSS-aligned assessment design at scale. Traditional ECD-based assessment design relies heavily on human effort~\cite{mislevy2012}. Regardless of the methods used, designing such assessment can be arduous, and the iterative nature of ECD activities further increases the time commitment~\cite{darling2013}. By distributing design decisions that rely on human expertise to different agents, this framework helps reduce the labor intensity of traditional approaches. Each agent was assigned different roles and work collaboratively, continuously maintaining the standards alignment and constraints through the item generation process. Specifically, this ECD-based MAS can align the PE with related DCI, SEP and CCC, increasing the efficiency for consistency check. Meanwhile, by embedding the evidence statement to the scenario and assessment generation, this system helps maintain evidence-task consistency and reduces the likelihood of task drift from the intended student evidence. Overall, this system can work as a supporting tool for teachers and assessment developers, helping to generating drafts, checking alignments, and accelerating the item generation. 

\subsection{Methodological Contributions: Operationalizing ECD through MAS }
This study also indicates that ECD-based MAS adds measurement foundations to traditional AIG. Prior research has shown that traditional AIG approaches leveraging LLMs are effective but lacks a solid educational and measurement grounding, with limited attention to validity, reliability, and pedagogical alignment~\cite{tan2024}. Many AIG approaches do the alignment check after the items are generated, which makes it difficult to prevent misalignment in earlier design stages. This framework integrates the claims-evidence-task principle across different agents, ensuring the assessment alignment is maintained as a structural constraint throughout the generation process, rather than a post-hoc verification step. In this way, the results show that MAS can serve as an important intermediary to connect assessment theory with automatic implementation, transforming ECD from a conceptual framework to an operational system. 

\subsection{Limitation and Future Directions}
This study has some limitations. While MAS shows potential in supporting assessment design, the results are unstable in coherence and multimodal performance, such as the inconsistency between text and images. The results further support the conclusion that the ECD-based MAS framework can work as a supporting tool, rather than an end-to-end developer for assessment design. Future studies can further explore when human experts should intervene and what key decisions they should make within a human-AI collaborative ECD based MAS. In this way, researchers can better understand how such human involvement can help improve system performance, particularly for multimodal representations and coherence. Future studies can also test the usability of this framework, examining teachers’ and assessment developers’ experiences and needs in real contexts and finally refining the practical value of this ECD-based MAS framework.

\bibliographystyle{splncs04}






\end{document}